\documentclass[reprint,amsmath,amssymb,aps,prl,showpacs]{revtex4-1}

\usepackage{graphicx}
\usepackage{dcolumn}
\usepackage{bm}
\usepackage{hyperref}

\begin{document}

\title{Electronic transport in graphene-based structures: an effective cross section approach}

\author{Andreas Uppstu$^1$}
\author{Karri Saloriutta$^1$}
\author{Ari Harju$^{1,2}$}
\author{Martti Puska$^1$}
\author{Antti-Pekka Jauho$^{1,3}$}

\affiliation{%
$^1$Department of Applied Physics, Aalto University School of Science, FI-02150 Espoo, Finland \\
$^2$Helsinki Institute of Physics, Aalto University School of Science, FI-02150 Espoo, Finland \\
$^3$Department~of~Micro~and~Nanotechnology,~DTU~Nanotech,~Technical~University~of~Denmark,~DK-2800~Kongens~Lyngby,~Denmark
}%

\date{\today}

\begin{abstract} We show that transport in low-dimensional carbon structures with finite concentrations of scatterers can be modeled by utilising scaling theory and effective cross sections. Our reults are based on large scale numerical simulations of carbon nanotubes and graphene nanoribbons, using a tight-binding model with parameters obtained from first principles electronic structure calculations.
As shown by a comprehensive statistical analysis, the scattering cross sections can be used to estimate the conductance of a quasi-1D system both in the Ohmic and localized regimes. They can be computed with good accuracy from the transmission functions of single defects, greatly reducing the computational cost and paving the way towards using first principles methods to evaluate the conductance of mesoscopic systems, consisting of millions of atoms.
\end{abstract}

\maketitle

Graphene, an effectively two-dimensional material consisting of a single sheet of carbon atoms, is regarded to be a potential candidate for a wide range of future electronic devices \cite{Geim}. In order to characterize phenomena affecting charge carrier transport in graphene-based systems, effective computational methods are required. Particularly important is the study of effects that induce a transport gap, turning graphene into a semiconductor.

The mechanisms behind experimentally measured transport gaps in graphene nanoribbons (GNRs) are currently actively debated. On one hand, it has been suggested that Coulomb blockade may significantly limit the conductance close to the Dirac point \cite{PhysRevB.84.073405, PhysRevLett.99.166803, Terres, PhysRevB.81.115409}, through barriers formed by either edge roughness \cite{PhysRevLett.99.166803} or charged impurities \cite{PhysRevB.81.115409}, although their role in limiting mobility has been questioned \cite{PhysRevLett.102.206603}. On the other hand, as the phase coherence length is very long in graphene \cite{Berger26052006}, also Anderson localization (AL) may induce a transport gap \cite{PhysRevB.78.161407, PhysRevB.79.075407, PhysRevB.79.235132, Gunlycke, PhysRevLett.100.036803, Querlioz}. This is supported by recent experimental results, which show a resistance growing exponentially with length in some GNRs~\cite{Xu}.

The low-energy band structure of graphene has two nonequivalent valleys, and due to the linear dispersion relation, intravalley scattering will result in antilocalization instead of localization \cite{PhysRevLett.99.106801, RevModPhys.82.2673}. Thus short-range disorder, causing intervalley scattering, needs to be present for AL to occur \cite{PhysRevLett.97.236801, PhysRevLett.97.236802, RevModPhys.82.2673}. In GNRs, scattering by imperfect edges may be one of the reasons behind this \cite{PhysRevB.78.161407, PhysRevB.79.075407, PhysRevB.79.235132, Gunlycke}. Additionally, Raman spectroscopy measurements of bulk graphene have revealed the presence of resonant scattering \cite{Ni}, which is another potential source of AL. The scattering may be due to hydrogen atoms~\cite{PhysRevB.82.081417} or hydrocarbons~\cite{PhysRevB.83.165402}.

In this Letter, we present numerical simulations of both Ohmic and localized systems, showing that point-like scatterers can effectively be described through a formalism based on defining scattering cross sections $\sigma(E)$ for the defects. The elastic mean free path $l_{el}(E)$ is related to the scattering cross sections $\sigma_i(E)$ of the different defect types and the corresponding defect densities $n_i$ via $l_{el}(E)=1/\sum_i n_i \sigma_i(E)$ (note that in two dimensions, $\sigma_i (E)$ is given in units of length). The scattering cross section approach provides a powerful means to estimate the conductance of a realistically sized GNR or carbon nanotube (CNT) with a finite number of point-like defects. We limit our discussion to short-range scatterers.

For a specific defect type, the scattering cross section may directly be obtained from the transmission function $T(E)$ of a conductor with one or several defects of the same type \cite{PhysRevB.81.125307, PhysRevLett.99.076803}. The conductance $G(E)$ is given by the Landauer formula $G(E)=(2e^2/h)T(E)$. In the Ohmic regime, the expression for $\sigma(E)$ in a system with $N$ defects is
\begin{equation}
\sigma(E)=W\frac{T_0(E)-\langle T(E) \rangle }{N \langle T(E) \rangle},  ~~~~~~~~~~ L \ll \xi(E),
\label{scs1}
\end{equation}
where $W$ is the width of the system (or the circumference of a CNT), $T_0(E)$ is the transmission function of the corresponding pristine conductor and $\langle T(E) \rangle$ is the ensemble average  over different defect positions and orientations. Eq.~(\ref{scs1}) is valid when the length $L$ of the conductor is much shorter than the localization length $\xi(E)$.

In the localized regime, the distribution of transmission values is not Gaussian, but rather log-normal. Thus, the typical transmission $T_{\rm typ}(E) \equiv \exp \langle \log T(E) \rangle$ is a meaningful scaling variable. In a single-mode conductor, it scales as $T_{\rm typ}(E)=\exp(-L/\xi(E))$~\cite{PhysRevB.22.3519}. Extending the scaling law to describe a multi-mode conductor, like a GNR or CNT, and treating the modes as conductors connected in parallel, we arrive at the expression
\begin{equation}
T_{\rm typ}(E)=T_0(E)\exp(-L/\xi(E)).
\label{asl}
\end{equation}
In systems belonging to the orthogonal Wigner-Dyson symmetry class (e.g. graphene with short-range disorder in the absence of a magnetic field), $\xi(E)$ is related to $l_{el}(E)$ and $\sigma(E)$ through~\cite{RevModPhys.69.731}
\begin{equation}
\xi(E) = \frac{(T_0(E)+1)l_{el}(E)}{2} = \frac{T_0(E)+1}{2n \sigma (E)}.
\label{xie}
\end{equation}
An expression for the scattering cross section, valid in the localized regime, is obtained by combining  Eqs.~(\ref{asl}) and (\ref{xie}):
\begin{equation}
\sigma(E)=\frac{W(T_0(E)+1) \langle \log(T_0(E)/T(E)) \rangle }{2N}, L \gg \xi(E).
\label{scs2}
\end{equation}

\begin{figure}
\vspace{0.3cm}
\includegraphics[width=\columnwidth]{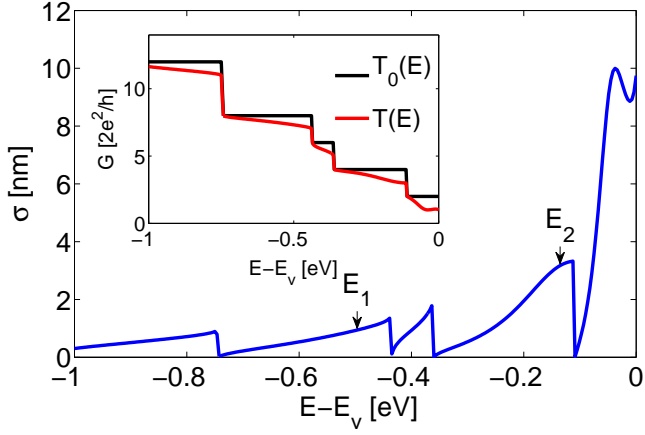}
\caption{(Color online) Scattering cross section $\sigma(E)$ of a monovacancy in a (40,0)-CNT, computed from the transmission function of a single defect (shown in the inset together with the transmission function of the corresponding defect-free conductor $T_0(E)$) using Eq.~(\ref{scs1}). The arrows show the locations of the energy values used in the statistical analysis of Fig.~\ref{fig2}. The energy is given with respect to the top of the valence band $E_v$.}
\label{fig1}
\end{figure}

\begin{figure}
\vspace{0.3cm}
~~\includegraphics[width=\columnwidth]{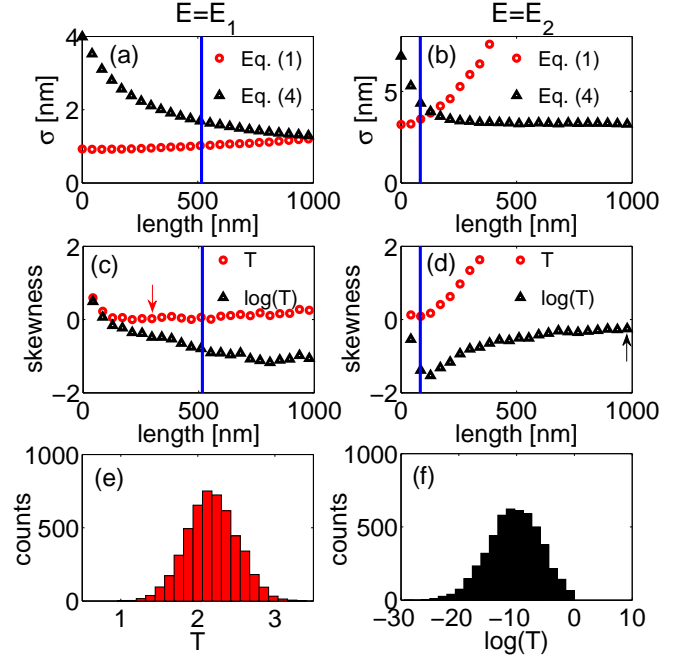}
\caption{(Color online) (a) and (b): Length dependence of the estimated scattering cross sections for a monovacancy in (40,0)-CNTs at the two energies shown in Fig.~\ref{fig1}. (c) and (d): Sample skewnesses of the distributions of $T$ and $\log (T)$. The defect density is $9.6\times 10^{-3}$ nm$^{-2}$, and the vertical bars indicate the localization lengths based on the scattering cross section shown in Fig.~\ref{fig1}. (e) Distribution of $T(E_1)$ at $L \approx 300$ nm, marked by the arrow in (c). (f) Distribution of $\log(T(E_2))$ at $L \approx 1000$ nm, marked by the arrow in (d).}
\label{fig2}
\end{figure}

We benchmark the formalism by comparing a single-defect based scattering cross section with transport results for systems with multiple defects. An unrelaxed monovacancy serves as a model defect, and to exclude edge effects, we have chosen a (40,0)-CNT, with a circumference of about 10 nm, as a model system. Unrelaxed monovacancies may also be used to model adsorbate hydrogen atoms \cite{PhysRevLett.107.016602}, as sp$^3$ hybridization creates a hole in the $\pi$ electron network, but also more detailed models for hydrogen adsorbates have been used \cite{PhysRevLett.105.056802, PhysRevLett.101.196803}.

We compute the transmission function by applying standard Green's function based methods to a  system formed by a device region containing the defects and two semi-infinite leads \cite{Datta}. The system is described by an orthogonal tight-binding (TB) model, with hopping values  obtained from Ref.~\cite{PhysRevB.51.12947}. A nearest-neighbor distance of 1.42 nm is assumed, and the values are scaled to obtain a nearest-neighbor hopping energy of $-2.7$ eV, in order to match our previous \emph{ab initio} results \cite{PhysRevB.81.245402}. All hopping values predicted to be smaller than 0.05 eV are set to zero, which in a pristine system means that hoppings to farther than third nearest neighbors are excluded. The scattering cross section of a monovacancy in a (40,0)-CNT, calculated using Eq.~(\ref{scs1}), is plotted in Fig.~\ref{fig1}, together with $T_0(E)$ and $T(E)$. As predicted by Fermi's golden rule, the van Hove singularities in the density of states (DOS) give rise to a greatly enhanced scattering rate near the band edges.

Next, we test how the single-defect scattering cross section compares against results for larger systems. Figs. 2 (a) and (b) show the estimated scattering cross sections given by Eqs. (\ref{scs1}) and (\ref{scs2}) as the length of the system is increased, keeping a constant defect density of $n=9.6\times 10^{-3}$ nm$^{-2}$. Each point has been obtained from an ensemble of 5000 different realizations of defect locations. The prediction of Eq.~(\ref{scs1}) grows exponentially as the length of the ribbon exceeds the expected localization length, whereas the prediction by Eq.~(\ref{scs2}) converges towards the estimate predicted by the transmission function of a single defect. Figs. \ref{fig2} (c) and (d) provide more insight into the behavior of the transmission values by showing the sample skewnesses~\footnote{The sample skewness is defined as the ratio $m_3/m_2^{3/2}$, where $m_3$ is the sample third central moment and $m_2$ the sample variance} of $T$ and $\log(T)$. At $E_1$, the distribution of the transmission values of very short conductors containing less than ten defects is skewed, but even in systems with very few defects the mean value of the distribution is very close to the prediction based on a single defect. As shown by Fig.~\ref{fig2}~(e), once the length of the conductor and thus the number of defects increase, the distribution of $T(E)$ becomes Gaussian-like, but as $\xi(E)$ is exceeded, the skewness of the distribution starts to grow rapidly. On the other hand, the skewness of the distribution of $\log(T)$ slowly approaches a value close to zero as the the localized domain is entered. However, even at roughly ten times the estimated localization length, the distribution is still slightly skewed, and as Fig.~\ref{fig2}~(f) shows, a relatively large fraction of the transmission values are of the order of $10^{-1}$, although $T_{\rm typ}$ is of the order of $10^{-5}$. The variance of the distribution shown in Fig.~\ref{fig2}~(f) equals 1.9 times the mean value of $-\log(T)$, which is close to the value of two predicted by random matrix theory~\cite{RevModPhys.69.731}.

\begin{figure}[t]
\vspace{0.3cm}
\includegraphics[width=\columnwidth]{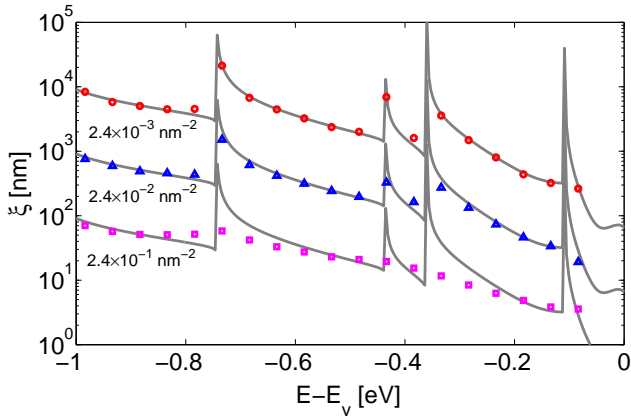}
\caption{(Color online) Predicted (solid lines) and calculated (markers) localization lengths  $\xi(E)$ for (40,0)-CNTs with monovacancies at three defect densities.}
\label{fig3}
\end{figure}

\begin{figure}[tb]
\vspace{0.3cm}
\includegraphics[width=\columnwidth]{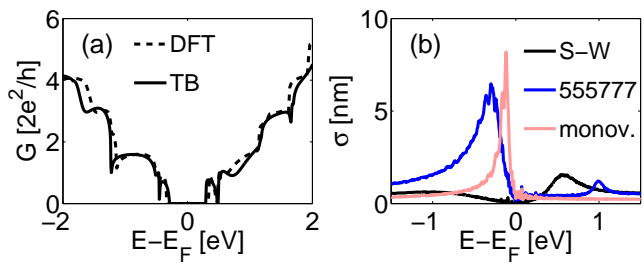}
\caption{(Color online) (a) Comparison between DFT and TB based conductances for a 16-AGNR with a single Stone-Wales defect. (b) Scattering cross sections for Stone-Wales defects, 555777 defects and monovacancies.}
\label{fig4}
\end{figure}

As a demonstration of the accuracy of our approach, we compare predicted and calculated localization lengths for wide ranges of energies and defect densities in Fig.~\ref{fig3}. The predictions are based on Eq.~(\ref{xie}) and the single-defect scattering cross section shown in Fig.~\ref{fig1}, whereas the calculated values have been obtained from systems ten times as long as the predicted localization lengths, using Eq.~(\ref{asl}). Each value corresponds to an ensemble of 200 defect realizations. At low defect density, $\xi(E)$ exhibits clear peaks, in accordance with Fig.~\ref{fig1}, but at higher densities the peaks corresponding to the DOS of a pristine CNT are smoothed out. However, even at the highest defect density shown, the scaling approach predicts the magnitude of $\xi(E)$ correctly. Thus we are able to predict properties of mesoscopic systems, only using the information from a single-defect calculation.

Based on Kubo-Greenwood (KG) simulations, it has been proposed that topological defects, like Stone-Wales defects and 555777 defects, exhibit fingerprint-like scattering properties \cite{PhysRevLett.106.046803}. A Stone-Wales defect is a metastable bond rotation, consisting of two pentagons and two heptagons embedded in the graphene lattice, whereas a 555777 defect is a relaxed form of a divacancy, consisting of three pentagons and three heptagons \cite{Banhart}. Such defects are expected to be found in irradiated graphene, where especially 555777 defects may occur in relatively high concentrations \cite{PhysRevLett.105.196102}. Fig.~\ref{fig4}~(a) compares transport results for a 16-atom wide armchair-edged GNR (AGNR), containing a single Stone-Wales defect, obtained both from a modified TB model and the density functional theory (DFT). The TB hopping parameters are obtained from the relaxed bond lengths around the defects, acquired from the \textsc{siesta} implementation~\cite{Ordejon1996, Soler2002} of DFT and the DFT transport results are from the \textsc{transiesta} code~\footnote{We use a double-$\zeta$ polarization basis set together with a 250 Ry mesh cutoff and the PBE-GGA functional for electron exchange and correlation.  Ionic coordinates are relaxed using a conjugate gradient algorithm until the force on each atom is within 0.02~eV/\AA.  Electron transmission is calculated with a single-$\zeta$ polarization basis set.}~\cite{Brandbyge2002}.

In Figure~\ref{fig4}~(b), we show bulk scattering cross sections for three defects, computed by applying $k$-space sampled periodic boundary conditions in the transverse direction~\footnote{We compute the bulk scattering cross sections from a system with a single defect in a 20 nm wide computational unit cell, by averaging over the different defect orientations in the armchair direction and by using 20 $k$-space points}. Although retaining the same general shape as the one shown in Fig.~\ref{fig1}, the scattering cross section of a monovacancy is now considerably smoother. At low defect densities of up to roughly $10^{-2}$ defects per atomic site, the scattering cross section for monovacancies fits both qualitatively and quantitatively the corresponding KG based mean free paths \cite{Laissardiere}. Also the scattering cross sections for Stone-Wales and 555777 defects agree fairly well with recent KG results \cite{PhysRevLett.106.046803}, although those are based on a somewhat differently parameterized TB model. When comparing against KG results, one has to remember that due to the phenomenon of minimum conductivity \cite{RevModPhys.83.407, RevModPhys.82.2673}, 2D graphene will not enter the localized regime. At very high densities, the defects lose their point-like nature, and the scattering cross section formalism breaks down. Additionally, a high density of defects will contribute to the transmission through an impurity band~\cite{PhysRevLett.105.056802}.

\begin{figure}[tb]
\vspace{0.3cm}
\includegraphics[width=\columnwidth]{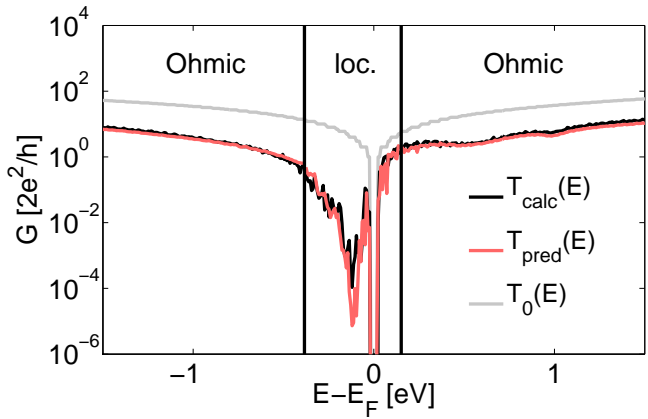}
\caption{(Color online) Predicted and calculated conductances of a realistically sized AGNR with 100 each of Stone-Wales defects, 555777 defects and monovacancies, as well as the conductance of a pristine conductor $T_0(E)$. In the Ohmic (localized) regimes, $T_{\rm pred}(E)$ refers to the estimate given by Eq.~(\ref{Tpredohmic}) (Eq.~(\ref{Tpredlocalized})) and $T_{\rm calc}(E)$ to the mean (typical) transmission.}
\label{fig5}
\end{figure}

We next test the predictive power of the scattering cross section approach. If a system contains several different types of defects, one can estimate the average and typical transmissions from
\begin{equation}
\langle T(E) \rangle=\frac{T_0(E)}{1+L\sum_i n_i \sigma_i(E)}, ~~~~~~~~~~ L \ll \xi(E)
\label{Tpredohmic}
\end{equation}
and
\begin{equation} T_{\rm typ}(E)=\frac{T_0(E)}{\exp \left[ \frac{2L}{T_0(E)+1} \sum_i n_i \sigma_i(E) \right] }, ~~~~ L \gg \xi(E).
\label{Tpredlocalized}
\end{equation}
We define the localized domain as the region where $T_{\rm typ}(E)$, given by Eq.~(\ref{Tpredlocalized}), is smaller than $\langle T(E)\rangle$, given by Eq.~(\ref{Tpredohmic}). Fig.~\ref{fig5} shows calculated mean and typical transmission functions for a 1.3 $\mu$m long and 30 nm wide AGNR with 100 each of monovacancies, Stone-Wales defects and 555777 defects, together with corresponding predictions obtained using Eqs. (\ref{Tpredohmic}) and (\ref{Tpredlocalized}). The calculated transmissions are based on an ensemble of 12 defect realizations, and the predictions on the bulk scattering cross sections shown in Fig.~\ref{fig4}~(b). As the results indicate, the mean or typical transmission in a realistically wide GNR or CNT can be estimated by only calculating the transmission functions for single defects in an edgeless system. The estimates are slightly lower than the actual mean and typical transmissions, which correspond to about 15\% lower values for the scattering cross sections than predicted from systems with single defects. The discrepancy may be caused by the anisotropy of the Stone-Wales and 555777 defects, as the single-defect scattering cross section corresponds to scattering only in the armchair direction.

In summary, we have presented numerical simulations showing that effective scattering cross sections for defects, combined with scaling theory, can be used to estimate the transport properties of graphene-based devices of sizes ranging from nano- to micrometers. In particular, their conductances can be predicted both in the Ohmic and strongly localized regimes. As the scattering cross section can be computed from small scale systems, possibilities to model systems beyond the reach of present-day methods are opened.

\begin{acknowledgments}
We acknowledge computational resources from CSC--IT Center for Science Ltd. and
the support by the Academy of Finland via the FiDiPro and CoE programs.
\end{acknowledgments}

%\bibliography{paper3}

%Merlin.mbs v4.21 2009-07-09.
%
\end{document}